\documentclass{PoS}
\usepackage{graphicx, amssymb, amsfonts,amsmath}
\usepackage{epsfig}
\newcommand{\be}{ \begin{eqnarray}}
\newcommand{\ee}{\end{eqnarray} }
\newcommand{\Tr}{{\rm Tr}} 
\newcommand{\Str}{{\rm Str}}

\def\MSbar{{\rm  \overline{\footnotesize MS\kern-0.05em}\kern0.05em}}
\newcommand{\Dlr}{\buildrel \leftrightarrow \over D\raise-1pt\hbox{}}
 \newcommand{\Dl}{\buildrel \leftarrow \over D\raise-1pt\hbox{}}
\newcommand{\Dr}{\buildrel \rightarrow \over D\raise-1pt\hbox{}}

\title{The microscopic Twisted Mass Dirac spectrum and the spectral determination of the LECs of Wilson $\chi$-PT}

\ShortTitle{Twisted Mass Dirac spectrum and LECs of Wilson $\chi$-PT}

  \vspace*{-0.3cm}    
 \author{Krzysztof Cichy\\
  Institut f\"ur Theoretische Physik, Goethe-Universit\"at Frankfurt,\\
Max-von-Laue-Str.~1, 60438 Frankfurt am Main, Germany \\
Faculty of Physics, Adam Mickiewicz University, Umultowska
85, 61-614 Pozna\'{n}, Poland\\
   E-mail: \email{kcichy@th.physik.uni-frankfurt.de}
    }       
  \vspace*{-0.3cm}    
\author{Elena Garcia-Ramos\\
John von Neumann Institute for Computing (NIC), DESY, Platanenallee 6, 15738 Zeuthen,
Germany\\
  E-mail: \email{elena.garcia.ramos@desy.de}
    }
  \vspace*{-0.5cm}    
  \vspace*{-0.5cm}    
\author{K. Splittorff\\
Discovery Center, The Niels Bohr Institute, University of Copenhagen, Blegdamsvej 17, DK-2100, Copenhagen \O, Denmark\\
   E-mail: \email{split@nbi.dk}
      }

   \vspace*{-0.3cm}

\author{\speaker{Savvas Zafeiropoulos}\\
   
Institut f\"ur Theoretische Physik, Goethe-Universit\"at Frankfurt,\\
Max-von-Laue-Str.~1, 60438 Frankfurt am Main, Germany\\
        E-mail: \email{zafeiro@th.physik.uni-frankfurt.de}}

\abstract{We present the comparison of the analytical microscopic spectral density for lattice QCD with $N_{\rm f}=2$ twisted mass fermions with the one obtained on the lattice utilizing configurations produced by the ETM collaboration.
We extract estimates for the chiral condensate as well as the low-energy constant $W_8$ of Wilson $\chi$-PT by employing spectral information of the Wilson Dirac operator with fixed index at finite volume.
}

\FullConference{The 33nd International Symposium on Lattice Field Theory\\
                 14-18 July, 2015\\
                 Kobe International Conference Center, Kobe, Japan}
\begin{document}

\section{Introduction}
Lattice QCD with Wilson twisted mass fermions has attracted a lot of attention since it enjoys many benefits such as automatic ${\cal
 O}(a)$ improvement of the action as well as of the matrix elements, absence of exceptional configurations and 
 fast dynamical simulations~\cite{Frezzotti:2003ni} in a solid theoretical framework. The main drawback is the existence of $\mathcal{O}(a^2)$ cutoff effects that break parity and vector flavor symmetry but they are of course suppressed close to the continuum limit. 
 Present day simulations are performed in the deep chiral regime with physical 
 value of the pion mass~\cite{physicalnf2}. However, due to the non-commutativity of the chiral and the continuum limit, Wilson fermions exhibit artificial phases with no continuum analogue. Such scenarios include the Aoki phase ~\cite{Aoki} that one reaches via a second order phase transition or the first order Sharpe-Singleton scenario \cite{SS}. One can study the phase diagram of twisted mass fermions in a lattice augmentation of Chiral Perturbation Theory ($\chi$-PT) particularly extended for Wilson fermions. In Wilson $\chi$-PT, one takes explicitly into account the lattice artifacts in the chiral expansion and ends with new terms which come with new Low Energy Constants (LECs). These LECs are particular to the lattice action but their knowledge is extremely important in order to gain access to the physical LECs such as the chiral condensate $\Sigma$ and the pion decay constant $F_{\pi}$. In addition to that, their relative strength and sign determines the potential of the effective theory and thus the phase structure ~\cite{Munster, Scorzato, realization, phdiagPRD}. For this reason, there is a lot of effort to determine them both analytically ~\cite{DSV, ADSV, KVZprl, Hansen, KVZprd, AokiBar} as well as numerically~\cite{Krzmixed, splittings, Fabio}.
 These approaches include a plethora of methods such as pion mass splittings ~\cite{splittings}, unitarity violations in a mixed action setup~\cite{Krzmixed} as well pion scattering in Wilson $\chi$-PT ~\cite{AokiBar}.  Here, we follow a different approach first applied in quenched studies ~\cite{DHS1, DWW, DHS2} where one matches analytical results from Wilson $\chi$-PT for a sector with fixed index $\nu$ of the Wilson Dirac operator with results obtained on the lattice. The notion of the index of the Dirac operator and the topological charge of the gauge configuration will be used interchangeably. 

\section{The theoretical background}
We make use of the analytical results for the microscopic spectral density for $N_{\rm f}=2$ derived in ~\cite{TMSV}. One starts from the supersymmetric extension of the chiral Lagrangian in the $\epsilon$-regime in a sector of fixed index $\nu$. In the microscopic power counting, $m\propto \epsilon^4$ and $a\propto \epsilon^2$ and terms up to $\mathcal{O}(\epsilon ^4)$ are taken into account.
In this regime, the partition function factorizes and one ends up with a unitary matrix integral which in our case reads 
\be
\label{ZSUSY}
{\cal Z}^\nu_{3|1}({\widehat{\cal Z}};\widehat{a})  & = & \int_{Gl(3|1)/U(1)} \hspace{-1.5mm} dU \
{\rm Sdet}(iU)^\nu
  e^{+\frac{i}{2}{\Str}(\widehat{{\cal Z}}[U+U^{-1}])
    + \widehat{a}^2{{\Str}(U^2+U^{-2})}},
\ee
where ${\widehat{\cal Z}}\equiv{\rm diag}(i\widehat{z}_t,-i\widehat{z}_t,\widehat{z},\widehat{z}')$ and the integration manifold is exactly the one that we encounter in continuum partially quenched $\chi$-PT calculations.  
This is the supersymmetric extension of the ordinary $N_{\rm f}=2$ chiral Lagrangian. In the supersymmetric (graded) method, one adds one species of a fermionic quark as well as one bosonic (ghost) quark with twisted masses $\widehat{z}$ and  $\widehat{z}'$, respectively. This allows for the computation of Green's functions when one differentiates with respect to the sources ($\widehat{z}$). Setting $\widehat{z}=\widehat{z}'$, one recovers the original $N_{\rm f}=2$ partition function. Here, we have introduced the rescaled variables $\widehat{a}=a \sqrt{W_8V/2}$ (with $W_8$ a new LEC parametrizing lattice artifacts) and $\widehat{z}_t=z_tV\Sigma$, where $z_t$ is the twisted mass, $a$ is the lattice spacing, $\Sigma$ is the chiral condensate and $V$ is the lattice volume. Note that the two other LECs which contribute at LO in $a^2$ have been ignored as a first approximation. We plan to derive the full analytical solution elsewhere. 
One computes Green's functions by differentiating ~(\ref{ZSUSY}) with respect to the sources and then subsequently the additional flavors have to be quenched. The resolvent reads 
\be
\label{Gsusy}
G^\nu_{3|1}(\widehat{z},\widehat{z}_t;\widehat{a})= \lim_{\widehat{z}'\to \widehat{z}} \frac{d}{d\widehat{z}} {\cal Z}^\nu_{3|1}(i\widehat{z}_t,-i\widehat{z}_t,\widehat{z},\widehat{z}';\widehat{a})
\ee
and its discontinuity yields the spectral density, $\rho^\nu_5(\widehat{\lambda}^5,\widehat{z}_t;\widehat{a})$, of the Hermitian Dirac operator $D_5\equiv\gamma_5 D$, with $\widehat{\lambda}^5=\lambda^5 V\Sigma$, which follows from
\be
\rho^\nu_5(\widehat{\lambda}^5,\widehat{z}_t;\widehat{a}) = \left \langle \sum_k \delta(\widehat{\lambda}^5_k-\widehat{\lambda}^5) \right \rangle_{N_{\rm f}=2} = \frac{1}{\pi}{\rm Im}[G^\nu_{3|1}(\widehat{z}=-\widehat{\lambda}^5,\widehat{z}_t;\widehat{a})]_{\epsilon\to0}.
\label{rho5def}
\ee
The explicit expression for the resolvent is given in ~\cite{TMSV}. Here, we have evaluated the corresponding integrals numerically in order to fit the lattice data.

\section{The computational setup and numerical results}
  In this study, we employ in the fermionic sector the twisted mass action~\cite{Frezzotti:2000nk} and in the gauge sector the Iwasaki improved
action~\cite{iwasaki}

\begin{equation}
   \label{eq:Sg}
   S_g =  \frac{\beta}{3}\sum_x\Biggl(  3.648 \sum_{\substack{
     \mu,\nu=1\\1\leq\mu<\nu}}^4\{1-\operatorname{Re}\Tr(U^{1\times1}_{x,\mu,\nu})\}\Bigr.
     \Bigl.-0.331\sum_{\substack{\mu,\nu=1\\\mu\neq\nu}}^4\{1
    -\operatorname{Re}\Tr(U^{1\times2}_{x,\mu,\nu})\}\Biggr)\, ,
 \end{equation}
 with $\beta$ the inverse bare coupling, $U^{1\times1}_{x,\mu,\nu}$ the plaquette and $U^{1\times2}_{x,\mu,\nu}$ rectangular $(1\times2)$ Wilson loops.
The twisted mass action for the light mass degenerate $u$, $d$ quarks reads
 \be S_l = a^4\sum_x \bar\chi_l(x) \left[D_W {+} m_{(0,l)} {+} i\mu_l\gamma_5\tau_3 \right]\chi_l(x)\,,
 \ee
 while for the heavy non-degenerate $s$ and $c$ quarks, we have
\begin{equation}
   \label{eq:sf}
  S_h\ =\ a^4\sum_x\left\{ \bar\chi_h(x)\left[ D[U] + m_{(0,h)} +
    i\mu_\sigma\gamma_5\tau_1 + \mu_\delta \tau_3 \right]\chi_h(x)\right\}\, ,
\end{equation}
where $D_W$ is the Wilson Dirac operator, $m_{(0,l)}$ is the untwisted bare light quark mass, $\mu_l$ is the bare twisted mass in the light sector,  $m_{(0,h)}$ is the untwisted bare quark mass for the heavy
doublet, $\mu_\sigma$ the bare twisted mass of the heavy doublet.
Note that the twist angle is this time along the $\tau_{1}$ direction -- and
$\mu_\delta$ the mass splitting along the $\tau_{3}$ direction.
The quark
 fields $\chi$ are in the so-called ``twisted basis'' obtained from the
 ``physical basis'' by a chiral transformation.

 We computed the index of the Dirac operator (topological charge of the gauge configurations) utilizing the Wilson Flow~\cite{ML} which is a cost effective method giving a solid definition of the topological charge and which does not require renormalization. Since we diagonalize the Dirac operator in sectors of fixed topological charge in order to have high statistics for given $\nu$, we employed a very long ensemble of the ETM collaboration which actually has $N_{\rm f}=2+1+1$. The heavy strange and charm quarks, with $a\mu_s=0.0158$ and $a\mu_c=0.2542$ respectively,  are completely quenched from a spectral viewpoint and bear no consequences in the comparison with the 
 $N_{\rm f}=2$ analytical results. The pion mass of the employed configurations is 390 MeV, $L\sim 2.5$ fm and $M_{\pi}L\sim 5$, which means that our results are already extrapolated to infinite volume from a practical point of view. 
Note that this is not an $\epsilon$-regime simulation where  
$M_{\pi}L\ll 1$, but this is not an issue, since the smallest Dirac eigenvalues can be in the $\epsilon$-regime. The scale below which eigenvalues are given by RMT is called the Thouless energy and for QCD it is $E_c=F_{\pi}^2/\Sigma L^2$ ~\cite{JamesJac}.
 We measured the topological charge of 5000 independent configurations and we diagonalized the ones with $|\nu|=0 , 1, 2, 3$. In  Fig.~1, we show the analytical results for $\rho^5$ for $|\nu|=0,1,2$, respectively, versus histograms of lattice data. One can observe that due to the large value of $\widehat{z}_t$, the results are very close to the quenched ones, while due to the large value of $\widehat{a}$, the structure of the former zero modes, that one would expect for $|\nu|=1, 2$, is completely dissolved. In order to extract the chiral condensate $\Sigma$ and additionally $W_8$, we performed constrained fits with $W_6=W_7=0$ in the individual topological sectors as well as a combined fit in the sectors with $|\nu|=0 , 1, 2$. Our results with their associated statistical errors are summarized in Table 1. The differences between the topological sectors are attributed to cut-off effects and they give a crude estimate of their size.
  We quote the result of the renormalized condensate having used the value of $Z_P$ in the $\overline{MS}$ scheme at 2 GeV given by the ETM collaboration to be $Z_P=0.509(4)$ \cite{italiani, francais}. 
In ~\cite{KrzSigma}, the continuum extrapolated value of $\Sigma$ that was quoted was equal to $\Sigma^{1/3}=290\pm 11$ MeV. We see that this value is very close to the one extracted from this study, which gives us confidence that cutoff effects are taken into account to some extent by the LO Wilson chiral Lagrangian which only includes $W_8$. The extracted value of $W_8$ is in complete agreement with ~\cite{Krzmixed} but differs by roughly a factor of 2 from the one determined in ~\cite{splittings}. This point requires extra clarification. Note that the values of the combined fits are still preliminary and they will be further scrutinized and addressed in an upcoming article.

\begin{figure}[p!]
\begin{center}
\vspace*{-0.35cm}
  \includegraphics[width=7cm, angle =-90]{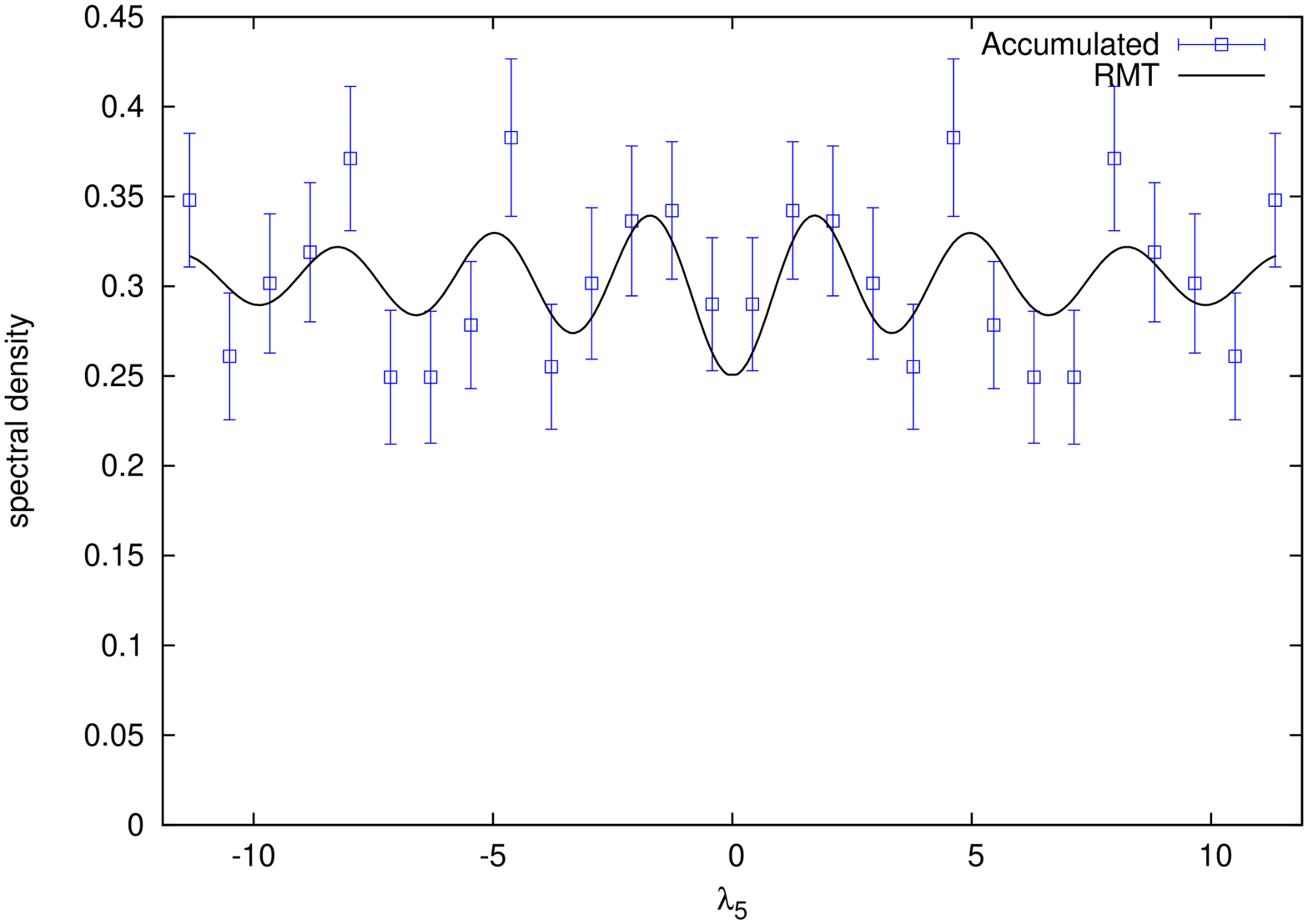}
\vspace*{-0.3 cm}
  \includegraphics[width=7cm,  angle =-90]{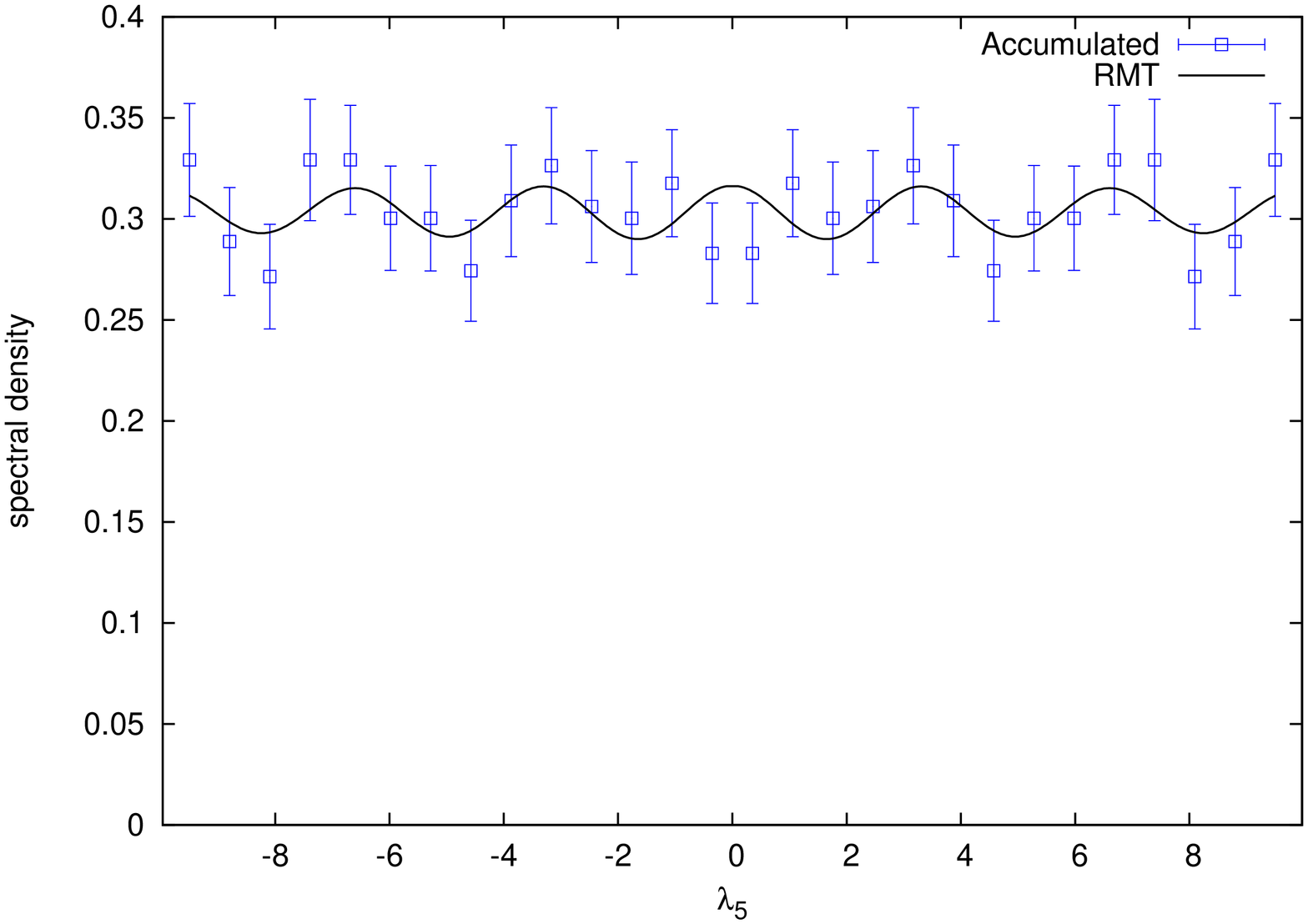}
\vspace*{-0.3 cm}
\includegraphics[width=7cm, angle =-90]{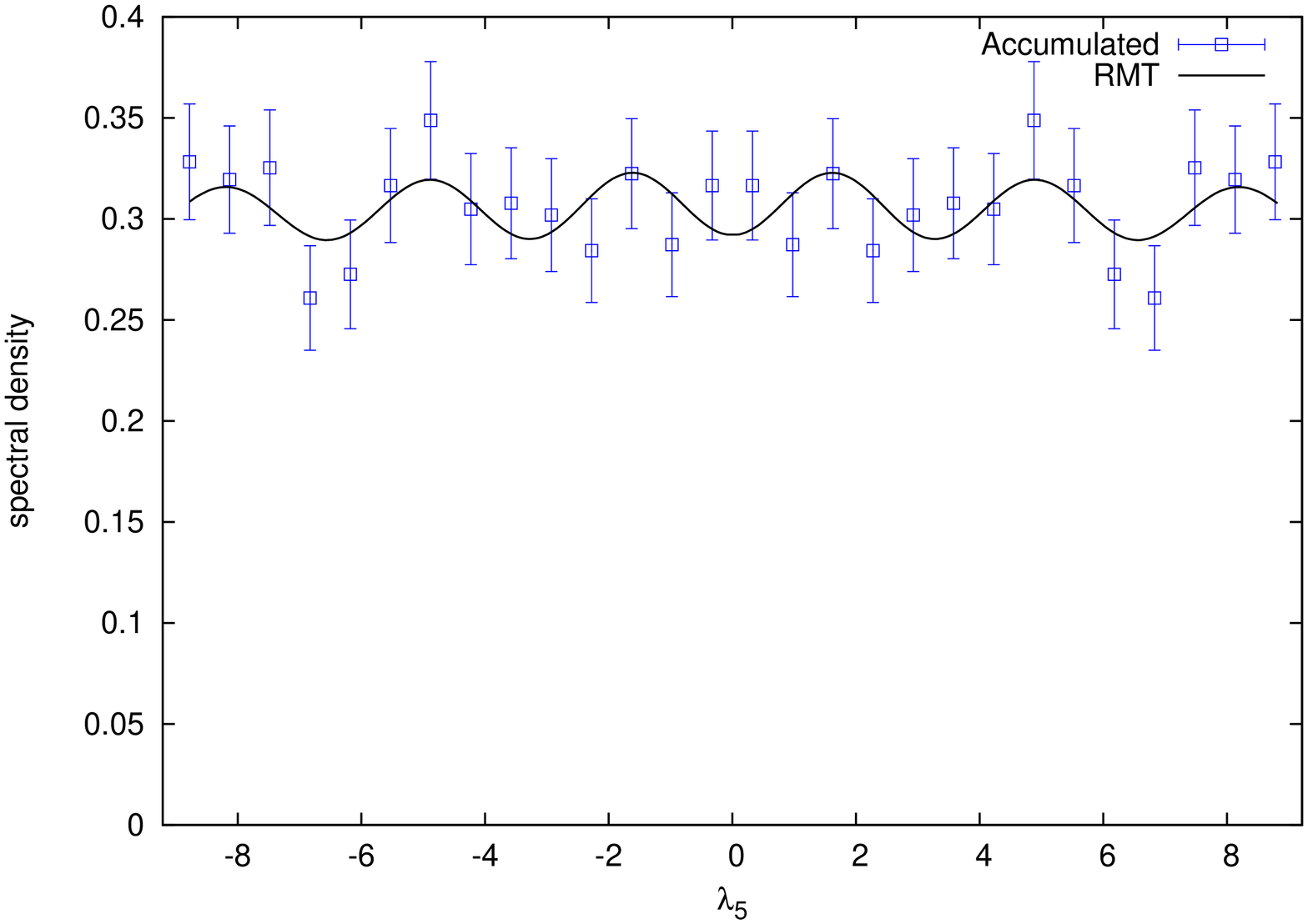}
\end{center}
   \vspace*{-0.65cm}
\caption{\label{fig1b} The microscopic spectral density of the Hermitian Twisted Mass Wilson Dirac operator. The solid curves comprise the analytical result derived in \cite{TMSV}, while the data points are the numerical results from a simulation on a $32^3\times 64$ lattice with
 $a=0.0815$ fm \cite{italiani} and $a \mu=0.0055$. In the top plot, we have the fitting results of the topological sector with  $\nu=0$ and fitting parameters $\widehat{z}_t=38.5$ and $\widehat{a}=0.715$. In the middle, we have the results for $|\nu|=1$ with fitting parameters $\widehat{z}_t=32.25$ and
$\widehat{a}=1.15$, while in the bottom, we have the results for $|\nu|=2$ with fitting parameters $\widehat{z}_t=31.71$ and $\widehat{a}=1.25$.
}
\end{figure}


\begin{table}[htbp]
\begin{center}
\begin{tabular}{ | l ||| c || c || c ||c|}
   \hline $|\nu |$ & 0 & 1 &2 &combined\\ \hline
   $\Sigma^{1/3}$\,\,[MeV] & 289.0(2.7) & 272.3(4.1)&270.8(6.8)&271.1(7.3) \\ \hline
   $W_8$ \,[$r_0^6W_0^2$] & 0.0021(12) & 0.0055(19)&0.0064(12)&0.0064(12) \\ \hline
 \end{tabular}
 \end{center}
 \caption{
 Extracted values for $\Sigma$ and $W_8$. \label{results}}
\end{table}

\section{Conclusions and Outlook}
In this study of unquenched twisted mass Dirac spectra, we took the first step towards the extraction of the LECs of Wilson $\chi$-PT from unquenched simulations. We determined the chiral condensate as well as $W_8$ by comparing analytical results from Wilson $\chi$-PT to numerical data from a lattice simulation.
All the errors quoted in this proceeding are at the moment only statistical errors which have been computed with the bootstrap method. Regarding the systematic errors, as it was already mentioned, finite size effects are expected to be under control due to the fact that $m_{\pi}L\sim 5$. Discretization errors are to a great extent taken into account by the fact that the chiral Lagrangian contains an $a^2$ term but still there can be contributions by the other two $a^2$ terms which have been neglected in Ref.~\cite{TMSV} and in our study, but also by higher order terms in the chiral expansion, proportional to e.g.
 $\mathcal{O}(a^4)$ (even though we expect the latter to be tiny for small values of $a$). One part of the calculation where discretization effects could creep in is in the computation of the topological charge. In order to minimize this effect, we have taken into account three different discretizations of the topological charge density. The first employs the plaquette definition and has cutoff effects of $\mathcal{O}(a^2)$, the second definition also includes a clover term and also has cutoff effects of $\mathcal{O}(a^2)$ and the last definition contains rectangular clover terms and has cutoff effects of $\mathcal{O}(a^4)$. The agreement among these methods for the given value of the lattice spacing was above $98\%$, see \cite{andreas} for a detailed analysis. These discretization errors could potentially lead to a misassignment of the topological charge and therefore the use of many different discretizations is deemed imperative in order to control this type of errors.
 This method is quite heavy from a computational viewpoint with the diagonalization of the Dirac operator being a pressing bottleneck. We computed the five lowest eigenvalues of the Hermitian Wilson Dirac operator by employing 1300 configurations and that had a cost of approximately 5 million core hours in the IDRIS/CNRS BG/Q in Paris. In order to improve on our results and to unveil the full dynamical features of the Wilson Dirac operator, one would need an even larger lattice volume in order to have more eigenvalues in the $\epsilon$-regime, an even smaller value of the rescaled twisted mass $\widehat{z}_t$ and besides that a smaller value of the rescaled lattice spacing $\widehat{a}$. A back of the envelope estimate for an ensemble of $48^3\times 96$ and $m_{\pi}=200$ MeV would give a total cost of around 75 milion core hours in a similar machine.

In an upcoming work, we plan to derive the full LO analytical solution including the LECs $W_6, W_7$, since they could potentially have a large effect in the eigenvalue density. It would be interesting to see how their inclusion will affect the extracted value of $\Sigma$ and $W_8$.

\vspace*{0.15cm}\noindent
{\bf Acknowledgments.}
This work was in part based on a variant of the ETM collaboration's  public lattice Quantum Chromodynamics  code \cite{tmlqcd1, tmlqcd2}.
We would like to thank Gregorio Herdoiza, Karl Jansen, Joyce C. Myers and Jac Verbaarschot for fruitful discussions and Andreas Athenodorou for providing us with the data of the topological charge.
This work was granted access to the HPC resources of CINES and IDRIS under the allocations 2013-052271 and 2014-052271 made by GENCI. We express our gratitude to the staff of this computing facility for their constant help. 
This work was supported by the the Humboldt Foundation (S.Z.) and the Sapere Aude program of
The Danish Council for Independent Research (K.S.) and by the Helmholtz International
Center for FAIR within the framework of the LOEWE program launched by the State of Hesse (K.C.).

\end{document}